\documentclass[aps,12pt,superscriptaddress,amsfonts,amssymb,amsmath]{revtex4}
\pdfoutput=1

\usepackage{graphicx}
\usepackage{epsfig}
\usepackage{makeidx}
\usepackage{epstopdf}
\usepackage{natbib}
\usepackage{xcolor}

\usepackage{color}

\begin{document}


\title{Network rewiring in the $r$-$K$ plane}

\author{M.L.\ Bertotti \footnote{Email address: marialetizia.bertotti@unibz.it}}
\author{G.\ Modanese \footnote{Email address: giovanni.modanese@unibz.it}}
\affiliation{Free University of Bozen-Bolzano \\ Faculty of Science and Technology \\ I-39100 Bolzano, Italy}

\linespread{0.9}

\begin{abstract}

We generate correlated scale-free networks in the configuration model through a new rewiring algorithm which allows to tune the Newman assortativity coefficient $r$ and the average degree of the nearest neighbors $K$ (in the range $-1\le r \le 1$, $K\ge \langle k \rangle$). At each attempted rewiring step, local variations $\Delta r$ and $\Delta K$ are computed and then the step is accepted according to a standard Metropolis probability $ \exp(\pm\Delta r/T)$, where $T$ is a variable temperature. We prove a general relation between $\Delta r$ and $\Delta K$, thus finding a connection between two variables which have very different definitions and topological meaning. We describe rewiring trajectories in the $r$-$K$ plane and explore the limits of maximally assortative and disassortative networks, including the case of small minimum degree ($k_{min} \ge 1$) which has previously not been considered. The size of the giant component and the entropy of the network are monitored in the rewiring. The average number of second neighbours in the branching approximation $\bar{z}_{2,B}$ is proven to be constant in the rewiring, and independent from the correlations for Markovian networks.  As a function of the degree, however, the number of second neighbors gives useful information on the network connectivity and is also monitored.

\end{abstract}

\maketitle

\section{Introduction}
\label{introduction}

Rewiring algorithms are often employed in network science to build ``synthetic networks'' for mathematical modelling of dynamics or diffusion processes \cite{newman2003mixing,boccaletti2006complex,cohen2010complex,pastor2015epidemic,barabasi2016network}. Usually, rewiring algorithms preserve the degree distribution of the network while changing the degree correlations and other topological features.
 
The ``configuration model'' \cite{newman2010networks} is a well established generalization of the random networks of Reny-Erd\"os which yields uncorrelated networks having a pre-assigned (typically scale-free) degree distribution. It is known, however, that assortative and disassortative correlations play an important role in dynamics and diffusion on networks \cite{newman2002assortative,van2010influence,d2012robustness,noldus2015assortativity,arcagni2017higher}. 
For this reason some algorithms have been devised, which are able to perform a degree-conserving rewiring while modifying the pair correlations in the direction of increasing assortativity or disassortativity. 

It is also possible to rewire the network in order to change its clustering coefficient (see \cite{alstott2019local} and refs.) or other metrics, but in this work we are focussing on assortativity and disassortativity as measured by the Newman coefficient $r$, and on the average nearest neighbors degree function $k_{nn}(k)$, or better on its network average $K \equiv \langle k_{nn}(k) \rangle = \sum_k P(k) k_{nn}(k)$. Here, $P(k)$ is the degree distribution, $P(h|k)$ denotes the conditional probability for a node of degree $k$ to be connected to a node
of degree $h$, and $k_{nn}(k) = \sum_k h P(h|k)$. 

The algorithm by Xulvi-Brunet and Sokolov \cite{xulvi2004reshuffling} is quite efficient for generating networks which are maximally assortative or maximally disassortative, or even have an intermediate $r$ coefficient, if a tunable return probability is inserted in the rewiring criterium. It does not allow, however, any direct control of the degree correlations $e_{jk}$ or $P(h|k)$, the $r$ coefficient or the $k_{nn}$ function. 

The rewiring method proposed by Newman \cite{newman2003mixing} allows in principle to generate ensembles of networks displaying, on average, any ``target'' two-point correlations assigned through an $e_{jk}$ matrix compatible with the given degree distribution. There exist several recipes for the construction of such matrices in the case of scale-free networks \cite{newman2003mixing,vazquez2003computational,bertotti2016bass}. 

We have recently proposed a new algorithm \cite{bertotti2019configuration} which is equivalent to the algorithm of \cite{xulvi2004reshuffling} when applied to maximally assortative or disassortative networks, but allows at each step to control the variation $\Delta r$ in the Newman coefficient, and therefore permits the introduction of a rewiring ``temperature'' $T$ in order to tune the return probability via a standard Metropolis update.
 
One of the aims of this work is to clarify the relations existing among these rewiring methods and the asymptotic constraints on maximally assortative and disassortative networks found by Menche et al.\ \cite{menche2010asymptotic}. Using our ``$\Delta r$-rewiring'' mentioned above we have been able to see the effects of extreme assortativity and disassortativity also in networks with many nodes of small degree. These were not considered by the authors of \cite{menche2010asymptotic}, who took as minimum degree a typical value $k_0=6$ and therefore found in general highly connected networks with a very large giant component. 

The study of complex networks is often motivated
by the interest for the dynamics of some diffusion problem on top of them. Clearly, if a mean-field approximation of the dynamics or diffusion on the network is sufficient for one's purposes, then the corresponding equations can be written, analysed or solved in terms of the excess-degree correlations $e_{jk}$ or in terms of the degree distribution $P(k)$ plus the conditional probabilities $P(h|k)$ (the so-called probabilistic or ``Markovian'' description of the network). If, on the other hand, a full realization of the network is nedeed (e.g., for simulations or stochastic modelling, or because one wants to take into account the effect of correlations beyond the second order), then several issues arise, for example:

\begin{itemize}

\item Is it possible to build any desired assortative or disassortative network, defined at the probabilistic level through a suitable ``theoretical'' $e_{jk}$ matrix, using a Newman rewiring, at least at the ensemble level? What is in this respect the role of the asymptotic constraints on $r$? Can the asymptotic constraints tell us in advance that a certain theoretical $e_{jk}$ is impossible to be implemented in a real network?

\item Among the networks obtained through a Xulvi-Brunet-Sokolov rewiring or our $\Delta r$ rewiring, will one find the desired assortative or disassortative network? If yes, with what accuracy is this possible, compared to the Newman rewiring?

\item How do the results (and their level of fluctuations and uncertainty) change if we modify the degree distribution, and especially the probability of the nodes with lowest degree? Will a giant component always be present? If the network is much fragmented, what are the consequences for diffusion processes?

\end{itemize}

As this work was progressing, we have gradually realized that such issues are really hard to solve in general terms. We did find some clues and the beginning of a path leading to partial answers, but we chose to leave most of these answers for a forthcoming publication. Still the efforts described in this work have already led to some useful spin-offs. In an attempt to obtain a better characterization of the rewiring process, we have represented the state of the network in an $r$-$K$ plane trying to use $K$ as a ``coordinate'' independent from $r$. This brought us first to a revision of the meaning and of the properties of $K$ (Sects.\ \ref{sezK0}, \ref{sezK1}, \ref{sezK2}) and then to establish a new general relation between the variations of $r$ and $K$ in a rewiring step (Sects.\ \ref{sezKvar}, \ref{sezKvar2}). This relation has been proven theoretically and verified numerically through the rewiring code. The code also allowed to guess some related properties of the quantity $\bar{z}_{2,B}$ (average number of second neighbors in the branching approximation), which have been proven theoretically in Sect.\ \ref{sezz2}.

In Sect.\ \ref{alg} we describe the main features of the rewiring code and the procedure for the calculation of the entropy of the generated networks. In Sect.\ \ref{results1} we describe the ``rewiring trajectories'' in the $r$-$K$ plane obtained for some different values of the scale-free exponent $\gamma$ ($\gamma=2.25, 2.5, 2.75, 3$) starting from uncorrelated networks and performing an assortative or disassortative rewiring at low temperature, i.e., with small return probability. For some of the cases a qualitative description is given of the ``super-assortative'' and ``super-disassortative'' asymptotic networks generated. Sect.\ \ref{results2} describes preliminary results of the assortative rewiring in equilibrium at variable temperature $T$, with a plot showing in the case $\gamma=2.5$, as a function of $T$, the values of the entropy $S$, of $\langle r \rangle$, $\langle K \rangle$ and the size of the giant component. Sect.\ \ref{concls} contains our conclusions.

\section{The function ``average degree of the first neighbors'' $K_N$}
\label{sezK0}

The quantity $\langle k_{nn}(k) \rangle_N$, introduced by 
Bogu\~n\'a  
et al.\ \cite{boguna2003epidemic}, is defined as 
\begin{equation}
\langle k_{nn}(k) \rangle_N = K_N =
\sum_{k=1}^n P(k) k_{nn}(k) \, . 
\end{equation}
We shall denote it for simplicity $K$, or $K_N$, since for a given type of network it depends only on its size, namely on the number $N$ of nodes (related in turn to the maximum degree $n$, for scale-free networks, through the Dorogovtsev-Mendes criterium as detailed in eq.\ (\ref{relaznN})). Since $k_{nn}(k)$ amounts to the average degree of the first neighbors of a node of degree $k$ and $P(k)$ is the probability that such a node is present, $K_N$ is the average degree of the first neighbors taken over the entire network, or better the average degree of the first neighbors of a randomly chosen node. Generally speaking, $K_N$ is strongly related to the diffusion properties of the network.

The definition above is probabilistic, and used for Markovian networks and for applications to mean-field equations on these networks. If we have a complete knowledge of the network, we can compute $K$ exactly just looking at the first neighbors of each node and computing a total average of their degrees (see Sect.\ \ref{alg}).

The authors of \cite{boguna2003epidemic} prove that as a function of $N$, $K_N$ is diverging when $N\to \infty$ in a scale-free network with exponent $2< \gamma \le 3$, for any kind of correlations (at least when the function $k_{nn}(k)$ has a certain form). This property is employed to conclude that in the ``thermodinamic'' limit of large $N$, phenomena of epidemic diffusion always propagate to the entire network, no matter how small the contagion probability is (``absence of epidemic threshold''). The intuitive reason is that although the average number of neighbors $\langle k \rangle$ tends to a constant for large $N$, the average degree of these neighbors tends to infinity; this means that each node is very close to a hub from which the epidemics can easily spread.

In the following two sub-sections we give some examples of computation of the function $K_N$  in Markovian networks, as an introduction to the results obtained through the rewiring of real networks.

\subsection{Uncorrelated networks}
\label{sezK1}

For an uncorrelated network we obtain for $K_N$ a simple expression. We have in this case
\[
P(h|k)=\frac{hP(h)}{\langle k \rangle} \, .
\]
Therefore $k_{nn}(k)$ does not depend on $k$:
\[
k_{nn}(k)=\frac{1}{\langle k \rangle} \sum_{h=1}^n h^2 P(h)=\frac{\langle k^2 \rangle}{\langle k \rangle}  \, ,
\]
and
\begin{equation}
K_N=\sum_{k=1}^n k_{nn}(k) P(k)=\frac{\langle k^2 \rangle}{\langle k \rangle}\ge \langle k \rangle  \, ,
\label{KNsc}
\end{equation}
where we have used the normalization condition $\sum_{k=1}^n P(k)=1$ and the last inequality is due to the fact that in general $\langle k^2 \rangle \ge \langle k \rangle^2$. 

The inequality $K_N \ge \langle k \rangle$ expresses the well-known property that in an uncorrelated network, from the point of view of one node looking at its first neighbors, on the average ``my friends have more friends than me'' (because their average degree is $K_N$ and my degree is $k$). As we shall see, this property is numerically confirmed also for correlated networks, at least in the scale-free case.

The dependence from $N$ in the expression (\ref{KNsc}) for $K_N$ arises as follows. First note that when we consider a finite network with maximal degree $n$, the normalization condition of $P(k)$ is $\sum_{k=1}^n P(k)=1$. Therefore the properly normalized degree distribution $P(k)$ for a scale-free network is
\[
P(k)=c_{n,\gamma}k^{-\gamma}, \ \ \ c_{n,\gamma}^{-1}=\sum_{k=1}^n k^{-\gamma}  \, .
\]
The quantities $\langle k \rangle$ and $\langle k^2 \rangle$, respectively equal to $\sum_{k=1}^n k P(k)$ and $\sum_{k=1}^n k^2 P(k)$, depend on $n$ through the factor $c_{n,\gamma}$ and the upper limit of the sum. However, in the limit of large $n$ the factor $c_{n,\gamma}$ tends to a constant and $\sum_{k=1}^n k P(k)$ is convergent for $2<\gamma \le 3$; the dependence of $K_N$ on $n$ comes from the divergent series $\sum_{k=1}^n k^2 P(k)$.

Approximating with an integral the dependence of the series on its upper limit, we obtain for large $n$  
\[
\langle k^2 \rangle \sim \int_1^n k^{-2-\gamma}dk \sim n^{-3-\gamma} \ \ \ {\rm for} \ 2<\gamma<3  \, ,
\]
and
\[
\langle k^2 \rangle \sim \ln(n) \ \ \ {\rm for} \ \gamma=3  \, .
\]
Of course, for $k$ close to 1 the integral is not a good approximation of the series; furthermore, if the sum starts from a value $k_{min}>1$, the factor $c_{n,\gamma}$ has a substantial dependence on $k_{min}$ (see Tab.\ \ref{table1}), as we shall later in some examples. Here, however, we are interested into the divergent dependence of $K$ on $n$.

In order to relate the maximum degree $n$ to the number of nodes $N$ we make recourse to the integral criterium of Dorogovtsev-Mendes \cite{dorogovtsev2002evolution}, which states that the probability to have in the network a node with degree in the range $(n,+\infty)$ must be equal to 1, implying
\[
\int_n^\infty c_{n,\gamma}k^{-\gamma}dk=\frac{1}{N}  \, .
\]
From this we obtain the known relation
\begin{equation}
\frac{\gamma-1}{c_{n,\gamma}}n^{\gamma-1}=N  \, .
\label{relaznN}
\end{equation}
An example of exact values of the various quantities involved is given in Tab.\ \ref{table1}.

\begin{table}
\begin{center}
\begin{tabular}{|c|c|c|c|c|c|c|c|c|c|} 
\toprule
$\ \gamma \ $ & $k_{min}$ &\ $N$\  &\ $n$\  &\ $c_{n,\gamma}$ \ &\  $\langle k \rangle$\ \ & $K$ (unc.) & $K$ (ass.) & $K$ (dis.) & $r_{dis}$ \nonumber \\
\hline
2.5 & 1 & 1000 & 93 & 1.34 & 1.79 & 7.43 & 4.61 & 8.52 & -0.088 \nonumber \\
2.5 & 4 & 1000 & 15 & 0.0896 & 6.23 & 7.39 & 6.81 & 20.1 & -0.140 \nonumber \\
2.75 & 1 & 1000 & 43 & 1.26 & 1.50 & 3.63 & 2.56 & 5.15 & -0.062\nonumber \\
2.75 & 4 & 1000 & 6 & 0.0413 & 4.64 & 4.77 & 4.70 & 16.0 & -0.120 \nonumber \\
\botrule
\end{tabular}	
\caption{An example of values of $K$ for Markovian networks with 1000 nodes in dependence on the scale-free exponent $\gamma$ and the minimum degree $k_{min}$, in the uncorrelated case, assortative case (Vazquez-Weigt recipe, eq.\ (\ref{vaz}), with $r=0.5$) and disassortive case (Porto-Weber recipe as employed in \cite{silva2019spectral}). }		
\label{table1}
\end{center}
\end{table}

\subsection{Correlated networks}
\label{sezK2}

If order to obtain $K_N$ for a correlated network we must compute numerically the sum over $k$ starting from an explicit expression for $k_{nn}(k)$, if known, or else expressing also $k_{nn}(k)$ as a sum, according to its definition $k_{nn}(k)=\sum_{h=1}^n hP(h|k)$. 

A simple formula which defines assortative correlations has been proposed by Vazquez and Weigt \cite{vazquez2003computational} and has been employed in \cite{nekovee2007theory} for diffusion studies. It is a linear combination of an uncorrelated term and a totally assortative term proportional to $\delta_{hk}$, namely
\begin{equation}
P(h|k)=(1-r)\frac{hP(h)}{\langle k \rangle}+r \delta_{hk}  \, ,
\label{vaz}
\end{equation}
where $r$ ranges from 0 to 1 and coincides with the Newman assortativity coefficient.

From this matrix one obtains
\[
k_{nn}(k)=(1-r)\frac{\langle k^2 \rangle}{\langle k \rangle}+r k  \, ,
\]
whence
\[
K_N=(1-r)\frac{\langle k^2 \rangle}{\langle k \rangle}+r \langle k \rangle  \, .
\]
It follows that for fixed $n$ (which also fixes $\langle k \rangle$ and $\langle k^2 \rangle$), when $r \to 1$ one has $K_N \to \langle k \rangle$, corresponding to the fact that in the case of extreme assortativity each node is only connected with other nodes having the same degree. We shall show in a forthcoming work, however, that for a real scale-free network this limit is purely hypothetical, because the function $k_{nn}(k)$ cannot increase linearly 
for large $k$, but eventually must decrease.

Further evaluations of $K_N$ as a function of $N$ for assortative networks, built using a different set of $P(h|k)$ matrices \cite{bertotti2016bass,bertotti2019evaluation}, and for disassortative networks will be given elsewhere.

In any case, for fixed $N$ the value of $K$ is quite useful to characterize the network and depends strongly on the type of correlations, on the scale-free exponent and on the minimum degree. A first example is given in Tab.\ \ref{table1} for Markovian networks. Then in Sects.\ \ref{results1}, \ref{results2} we will investigate the behavior of $K$ for real rewired networks. An exact direct calculation of $K$ from the list of links of a real network can be efficiently implemented and also compared with the average value of $K$ obtained from the $k_{nn}$ function.

\begin{figure}
\begin{center}
\includegraphics[width=13cm,height=7cm]{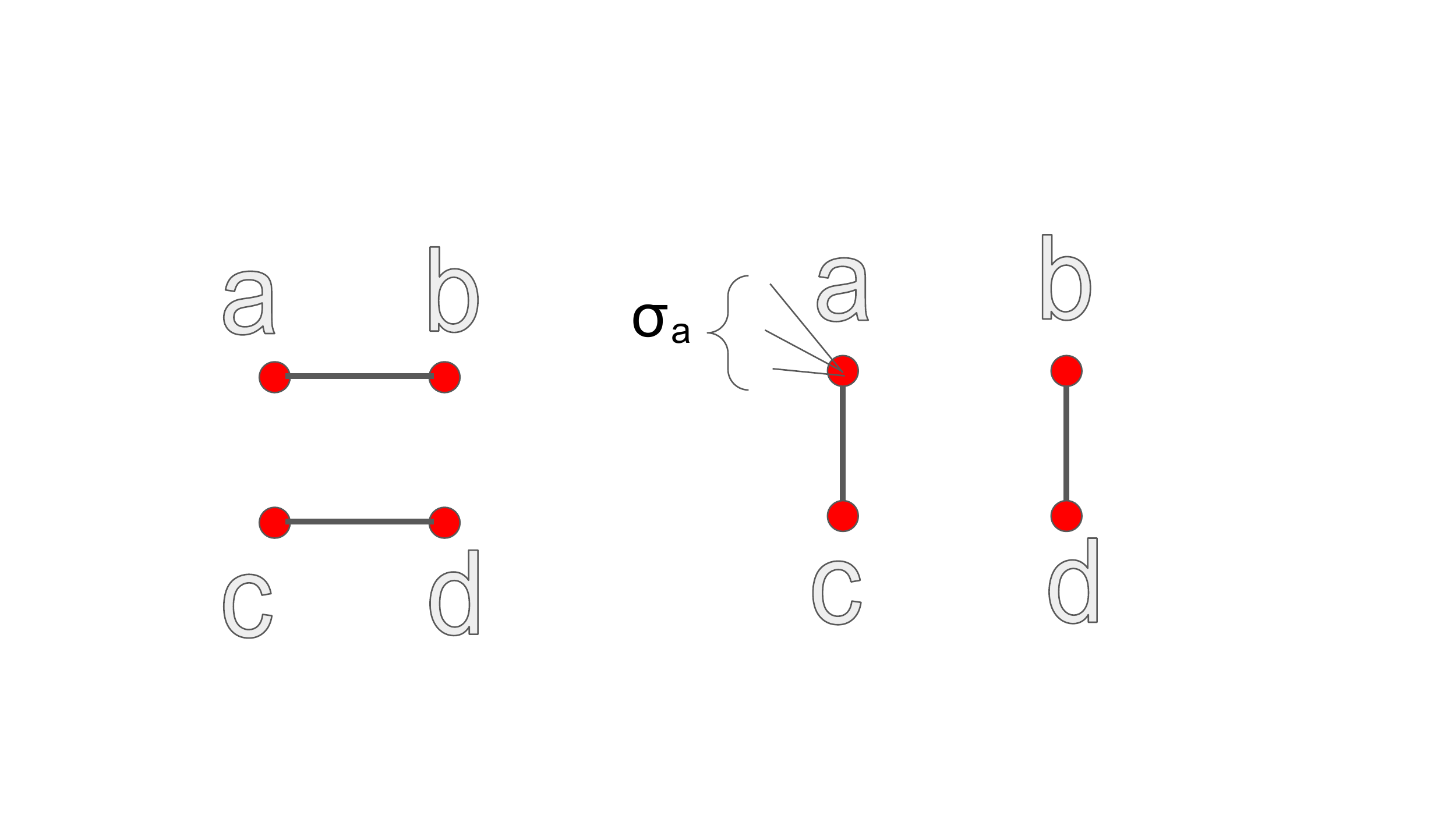}
\caption{Scheme of a rewiring which involves the nodes $a,b,c,d$ and leaves their degrees $A,B,C,D$ unchanged. Before the rewiring the links are $(a,b)$, $(c,d)$. After the rewiring the links are $(a,c)$, $(b,d)$. Each node has in general other neighbors, not depicted; the sum of the degrees of the neighbors of $a$ not involved in the rewiring is denoted in the text as $\sigma_a$, and similarly for $b,c,d$. 
} 
\label{fig-rew}
\end{center}  
\end{figure}

\subsection{Local variation of $K$}
\label{sezKvar}

The variation of $K$ in a rewiring step is obtained using its definition as the average degree of the first neighbors of each node, averaged over the whole network. Let $A,B,C,D$ denote respectively the degrees of the nodes $a,b,c,d$ involved in the rewiring. Let $\sigma_a$ denote the sum of the degrees of the first neighbors of the node $a$ which are \emph{not} involved in the rewiring, and similarly define $\sigma_b$, $\sigma_c$, $\sigma_d$. Before the rewiring the averages $s_a, s_b, s_c, s_d$ of the degrees of the first neighbors of $a,b,c,d$ are
\[
s_a=\frac{\sigma_a+B}{A}; \ \ s_b=\frac{\sigma_b+A}{B}; \ \ s_c=\frac{\sigma_c+D}{C}; \ \ s_d=\frac{\sigma_d+C}{D}  \, .
\]
After the rewiring, these quantities become
\[
s_a=\frac{\sigma_a+C}{A}; \ \ s_b=\frac{\sigma_b+D}{B}; \ \ s_c=\frac{\sigma_c+A}{C}; \ \ s_d=\frac{\sigma_d+B}{D}  \, .
\]
Therefore the change in the total average $K$ is
\begin{equation}
\Delta K= \frac{1}{N} \left[ \left( \frac{C}{A}+\frac{D}{B}+\frac{A}{C}+\frac{B}{D} \right)-
\left( \frac{B}{A}+\frac{A}{B}+\frac{D}{C}+\frac{C}{D} \right) \right]  \, .
\label{deltaK1}
\end{equation}

\subsection{Relation between the local variations of $K$ and $r$}
\label{sezKvar2}

In our previous work \cite{bertotti2019configuration} we found an expression for the local variation of the Newman coefficient in a rewiring of the same kind as in Fig.\ \ref{fig-rew}. The variation is given by
\begin{equation}
\Delta r=\frac{2(-AB-CD+AC+BD)}{L\sigma^2_q}  \, ,
\label{deltar}
\end{equation}
where $L$ is the number of links in the network and $\sigma^2_q$ is the denominator of the fraction which defines $r$, namely
\begin{equation}
r=\frac{1}{\sigma^2_q} \sum_{j,k=0}^{n-1}jk(e_{jk}-q_jq_k)  \, ,
\label{formular}
\end{equation}
\begin{equation}
\sigma^2_q= \sum_{k=0}^{n-1}k^2q_k-\left( \sum_{k=0}^{n-1}kq_k \right)^2  \, .
\label{formulasigma}
\end{equation}

Here $e_{jk}$ is the probability to find in the network a link between nodes with excess-degrees $j$ and $k$ and $q_k=\sum_{j=0}^{n-1}e_{jk}$ is the excess-degree distribution. 
Note that $q_k$ and $\sigma^2_q$ depend on the degree distribution but not on the correlations. Also note that in \cite{bertotti2019configuration} a slightly different notation is used, in which $A,B,C,D$ denote directly the excess-degrees. However, since
\[
-(A-1)(B-1)-(C-1)(D-1)+(A-1)(C-1)+(B-1)(D-1)=-AB-CD+AC+BD  \, ,
\]
we can safely use the expression (\ref{deltar}) according to the conventions of this paper, where $A,B,C,D$ are the degrees and not the excess-degrees.

After some algebraic manipulations it is possible to express the variation $\Delta K$ in eq.\ (\ref{deltaK1}) in terms of $\Delta r$, thus establishing a relation between two quantities which have very different definitions and topological meaning. We find
\begin{equation}
\Delta K=-\frac{L\sigma^2_q}{2N} \frac{(AD+BC)}{ABCD} \Delta r  \, .
\label{deltaKbis}
\end{equation}
This holds for each rewiring. The factor $-L\sigma^2_q/(2N)$ is fixed for a given degree distribution, while the factor $(AD+BC)/(ABCD)$ clearly depends on the nodes involved; we only know a priori that it is always positive, and as a consequence $\Delta K$ is always opposite to $\Delta r$ and  $\Delta K=0$ if and only if $\Delta r=0$. For the rewiring trajectories described in Sect.\ \ref{results1} the ratio $\Delta K/\Delta r$, averaged over many rewirings, turns out to be approximately constant for a given degree distribution; see data in Tab.\ \ref{table2}.

\begin{table}
\begin{center}
\begin{tabular}{|c|c|c|c|c|c|} 
\toprule
$\ \gamma \ $ & $N$ &\ $L$\  &\  $\sigma^2_q$\ & $\frac{L\sigma}{2N}$ & $\langle \frac{AD+BC}{ABCD} \rangle_{rew}$ \nonumber \\
\hline
2.25 & 715 & 798 & 193 & 108 & 0.24  \nonumber \\
2.5 & 715 & 630 & 81.6 & 36 & 0.31 \nonumber \\
2.75 & 715 & 536 & 30.2 & 11.3 & 0.35 \nonumber \\
3 & 715 & 481 & 12.4 & 4.2 & 0.43 \nonumber \\
\botrule
\end{tabular}	
\caption{An example of the numerical factors involved in the relation (\ref{deltaKbis}) between $\Delta K$ and $\Delta r$, for scale-free networks with a fixed number of nodes. These factors explain why the rewiring trajectories in the $r$-$K$ plane have different slopes.
(See Figs.\ \ref{traiettASS}, \ref{traiettDIS} and the description in Sect.\ \ref{results1}.)
$N$ is the number of nodes, $L$ the number of links. $\sigma^2_q$ is the variance of the excess-degree distribution (eq.\ \ref{formulasigma}). 
$\langle (AD+BC)/(ABCD) \rangle_{rew}$ denotes the average along an assortative rewiring trajectory with low temperature $T=10^{-6}$.
The hubs of the networks are defined by the cumulative probability method. 
}		
\label{table2}
\end{center}
\end{table}

\section{The average number of second neighbors $\bar{z}_{2,B}$}
\label{sezz2}

The condition for the existence of a giant component in an uncorrelated network with arbitrary degree distribution has been first found by Molloy and Reed \cite{molloy1995critical} and is expressed by the inequality
\begin{equation}
\langle k^2 \rangle -2 \langle k \rangle>0  \, .
\label{eqMR}
\end{equation}
Later the same condition has been proven by Newman, Strogatz and Watts with the method of the generating functions, which also allows to find the size of the giant component \cite{newman2001random}. In Ref.\ \cite{newman2001random} the inequality (\ref{eqMR}) is reformulated in an intuitive way by stating that the giant component exists when 
$\bar{z}_2 > \langle k \rangle$, where $\bar{z}_2$ is the average number of second neighbors and $\langle k \rangle$ (the average degree) can also be interpreted as the average number of first neighbors. A crucial underlying assumption is that the network is locally a branching structure; moreover, being the network uncorrelated, one supposes that there are no preferences in linking behavior depending on the node degrees and that therefore it makes sense to consider total averages like $\bar{z}_2$ and $ \langle k \rangle$. 

We are going now to define a quantity which is closely related to $\bar{z}_2$ and we will show 
that starting from the intuitive ``percolation'' condition $\bar{z}_2> \langle k \rangle$, condition (\ref{eqMR}) can be immediately obtained without using the generating functions.

We call this quantity ``average number of second neighbors in the branching approximation'' and denote it by $\bar{z}_{2,B}$. It is a network average like $\bar{z}_2$, but includes by definition multiple counting in the case of shared second neighbors. More precisely, if one node $a_0$ has a second neighbor $b$ in common with other $h$ nodes $a_1,\ldots,a_h$, then $b$ is counted $h+1$ times in the average $\bar{z}_{2,B}$.
The two quantities $\bar{z}_2$ and $\bar{z}_{2,B}$ coincide if the network is a pure branching structure, without nodes that have second neighbors in common with other nodes.

According to this definition, $\bar{z}_{2,B}$ can be obtained as
\begin{equation}
\bar{z}_{2,B}=\sum_k P(k) k [k_{nn}(k)-1]  \, ,
\label{eq1}
\end{equation} 
because the probability for a node to have degree $k$ is $P(k)$, the node has $k$ first neighbors and their average degree is $k_{nn}(k)$.

Let us compute $\bar{z}_{2,B}$ for an uncorrelated network:
\begin{equation}
\bar{z}_{2,B}=\sum_k P(k) k [k_{nn}(k)-1]=\sum_k P(k) k \left[ \frac{\langle k^2 \rangle}{\langle k \rangle}-1 \right]=
\end{equation}
\begin{equation}
=\left[ \frac{\langle k^2 \rangle}{\langle k \rangle}-1 \right]\langle k \rangle=\langle k^2 \rangle-\langle k \rangle \ \ \ {\rm (uncorrelated \ network)}  \, .
\label{z2sc}
\end{equation}
This expression gives the Molloy-Reed condition if we require $\bar{z}_2>\langle k \rangle$ and admit that the network is a locally branching structure such that $\bar{z}_2 \simeq \bar{z}_{2,B}$.

The expression (\ref{eq1}) for $\bar{z}_{2,B}$ is interesting in itself and we have used our code for the configuration model with rewiring in order to test it. The code generates a list, called the ``\texttt{Friends}'' list, of the first neighbors of each node. The list is updated and used in many parts of the program, for instance after the first wiring of the stubs, in order to check that their degrees match the prescribed degree distribution. It is also used at the end of the rewiring cycles, in order to find the giant component of the final network, and possibly for the numerical solution of diffusion equations in first or second closure approximation. It is straightforward to use the \texttt{Friends} list also to obtain the number of second neighbors of each node, because the degrees of the nodes do not change in the rewiring and are stored in a vector ``\texttt{Degrees[i]}'', with $i=1,\ldots,N$, fixed from the degree distribution before the wiring. The contribution to $\bar{z}_{2,B}$ from each node is obtained as the sum of (\emph{degree of each friend - 1}). The total network average $\bar{z}_{2,B}$ is the sum of the contributions of all nodes, divided by $N$. One can check that the exact value obtained in this way is well approximated by the probabilistic value (\ref{eq1}).

Somewhat unexpectedly, the exact value of $\bar{z}_{2,B}$ obtained is accurately reproduced in each simulation with the same degree distribution, signaling that it is not affected by the rewiring. In fact, the following two properties hold, which are not difficult to prove but cannot be found in the literature, to the best of our knowledge.

\ 

\textbf{Property 1 of $\bar{z}_{2,B}$:} for Markovian networks which satisfy the Network Closure Condition $hP(k|h)P(h)=kP(h|k)P(k)$, $\bar{z}_{2,B}$ does not depend on the correlations but only on the degree distribution, and it is equal to the value $\langle k^2 \rangle-\langle k \rangle$ obtained for an uncorrelated network having that degree distribution (eq.\ (\ref{z2sc})).

\ 

In fact the first term on the r.h.s.\ 
in eq.\ (\ref{eq1}), namely 
$\sum_k P(k) k k_{nn}(k)$, is equal to $\langle k^2 \rangle$, as already noted in \cite{boguna2003epidemic}:

\[\sum_k P(k) k k_{nn}(k)=\sum_k P(k) k \sum_h h P(h|k)=\]
\[=\sum_k  \sum_h P(k) kh P(h|k)=\sum_k  \sum_h hh P(k|h)P(h)=\]
\[=\sum_h h^2P(h) \sum_k P(k|h)=\langle k^2 \rangle\]
(because $\sum_k P(k|h)=1$). The second term is equal to $\langle k \rangle$, which is fixed if the degree distribution is fixed.

\ 

\textbf{Property 2 of $\bar{z}_{2,B}$:} $\bar{z}_{2,B}$ does not change in a binary rewiring which preserves the degree distribution. This is a direct consequence of the definition of the rewiring (see Fig.\ \ref{fig-rew}). Let $\sigma_a$ denote, as before, the sum of the degrees of the first neighbors of the node $a$ which are \emph{not} involved in the rewiring, and similarly define $\sigma_b$, $\sigma_c$, $\sigma_d$. Let $A,B,C,D$ denote the degrees of the nodes. The total number of second neighbors of the four nodes involved in the rewiring is equal, before the rewiring, to  
\[
z_{2,B}^{before}(a,b,c,d)=
(\sigma_a-A+B-1)+(\sigma_b-B+A-1)+(\sigma_C-C+D-1)+(\sigma_D-D+C-1)  \, .
\]
After the rewiring we have
\[
z_{2,B}^{after}(a,b,c,d)=
(\sigma_a-A+C-1)+(\sigma_b-B+D-1)+(\sigma_C-C+A-1)+(\sigma_D-D+B-1)  \, ,
\]
and the two quantities are equal.

These properties offer a strong indication for the absence of epidemic threshold in correlated large scale-free networks with $2<\gamma \le 3$. In fact, in this range of $\gamma$, $\bar{z}_{2,B}$ diverges when $N\to \infty$. Denoting by $\lambda$ the contagion probability for one single contact, consider an infected node randomly chosen, thus with average degree $\langle k \rangle$. The probability that the node infects one of its neighbors is $\langle k \rangle \lambda$, which tends to zero if $\lambda$ is very small. However, the probability that the node infects one of its second neighbors is (for a locally branching structure) equal to $\bar{z}_{2,B} \lambda^2$, and this quantity can stay finite even as $\lambda \to 0$, because of the divergence of $\bar{z}_{2,B}$, independently from the degree correlations.

\begin{figure}
\begin{center}
\includegraphics[width=18cm,height=10.5cm]{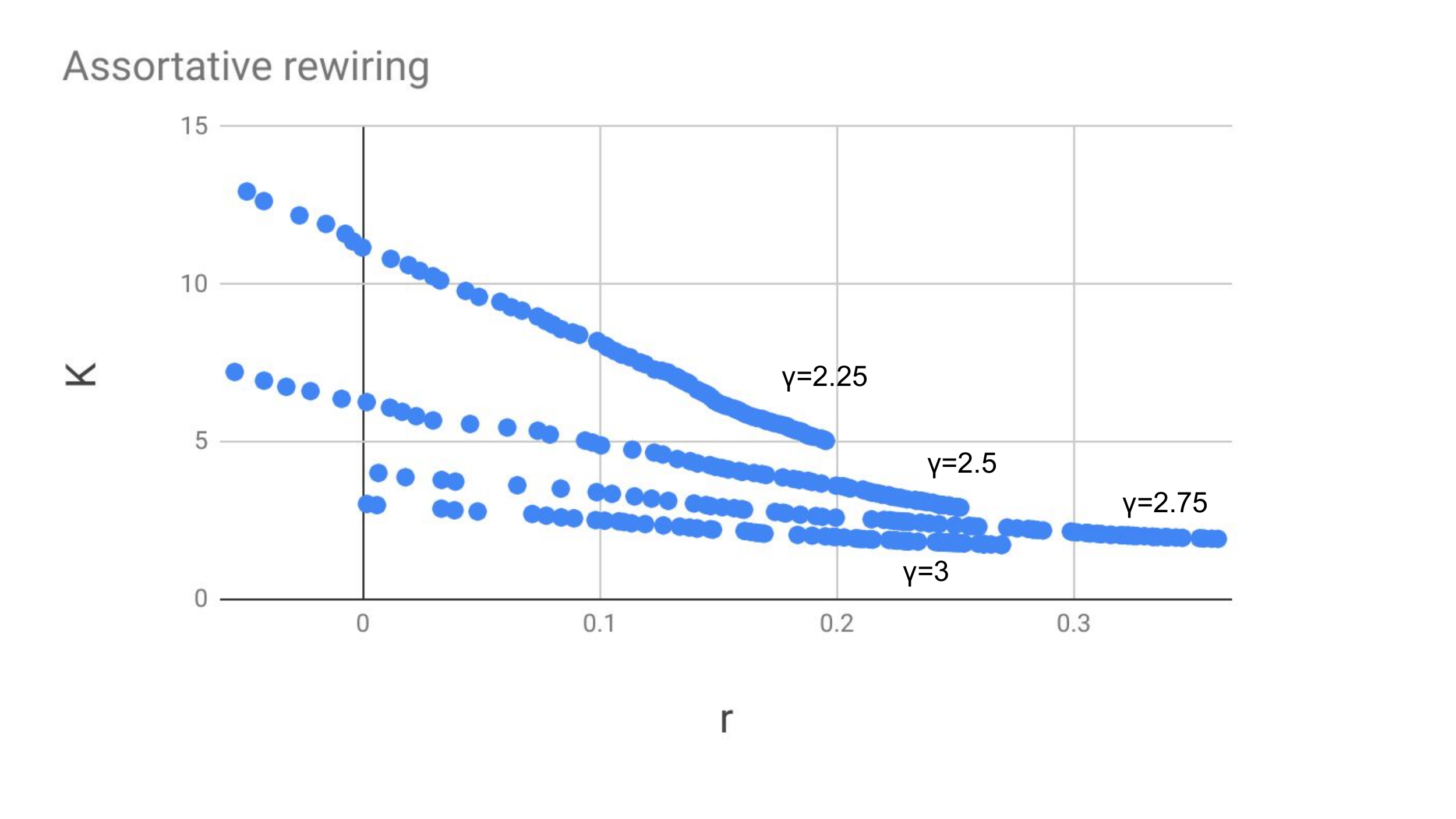}
\caption{Trajectories of assortative rewiring in the $r$-$K$ plane. Each trajectory begins from the left (initial network obtained by uncorrelated rewiring) and converges to a maximally assortative network on the right. Between two dots on the same trajectory there are 100 rewiring steps with return probability $\exp(-\Delta r/T)$, $T=10^{-6}$. The number of nodes is $N=1500$.
} 
\label{traiettASS}
\end{center}  
\end{figure}

\section{The rewiring algorithm}
\label{alg}

In the first part of the algorithm, the ``wiring'' part, we generate a series of $N$ ``stubs'' with given degree distribution, like in any implementation of the configuration model. The exact procedure for assigning the degrees to the hubs has two possible alternatives and has been described in \cite{bertotti2019configuration}. In the first alternative (``cumulative hubs''), the probability of stubs for which $P(k)\cdot N<1$ is cumulated with increasing $k$ until it exceeds 1, at which point the hub is created and the accumulation starts again. In the second alternative (``random hubs''), hubs of degree $k$ are created entirely at random with probability $P(k)\cdot N$.

Also for a description of the linking of the stubs the reader is referred to \cite{bertotti2019configuration}. Details of this procedure are little relevant because the extensive rewiring that follows cancels any memory of the initial wiring scheme.

When we perform a rewiring step, we choose at random two links in the current list of links describing the network, say $(a,b)$ and $(c,d)$ (nodes are identified by a sequential number in the range $1, \ldots ,N$, so $a,b,c,d$ denote this number). With a probability of 50\% we exchange $a$ and $b$, to avoid any asymmetry, and then we build the new links $(a,c)$, $(b,d)$. In order to avoid the formation of loops, the rewiring is not performed if $a=c$ or $b=d$. The formation of multi-links (more than one link between the same two nodes) is avoided through a check of the adjacency matrix $[A_{ab}]$, which is computed from the list of links before the rewiring cycle and updated after each rewiring according to the formulas
\[
A_{ab}\to A_{ab}-1; \ \ A_{cd}\to A_{cd}-1; \ \ A_{ac}\to A_{ac}+1; \ \ A_{bd}\to A_{bd}+1  \, .
\]
(plus the symmetrical variations for $A_{ba}$ etc.).

The knowledge of the adjacency matrix also allows to compute the Shannon entropy of the network (see \cite{johnson2010entropic} and refs.) by ensemble-averaging $A$ over sub-cycles. For instance, consider a typical rewiring cycle with 100 sub-cycles of $10^4$ steps each, for a network with $N=10^3$. The values of $A_{ab}$, with $a,b=1,\ldots,N$ at the end of each sub-cycle are averaged to compute $S$ according to the formula
\begin{equation}
S=-\sum_{a,b=1}^N \bar{A}_{ab} \ln \bar{A}_{ab}+(1-\bar{A}_{ab})\ln (1-\bar{A}_{ab});
\ \ \ \bar{A}_{ab}=\langle A_{ab} \rangle_{sub-cycles}  \, .
\label{entropia}
\end{equation}

The number of sub-cycles is increased until the result stabilizes. The averages of $r$ and $K$ are also computed in the same way. A long rewiring process of this kind is used to compute $S$ and other quantities as functions of the temperature; see results in Sect.\ \ref{results2}.

\begin{figure}
\begin{center}
\includegraphics[width=10cm,height=10cm]{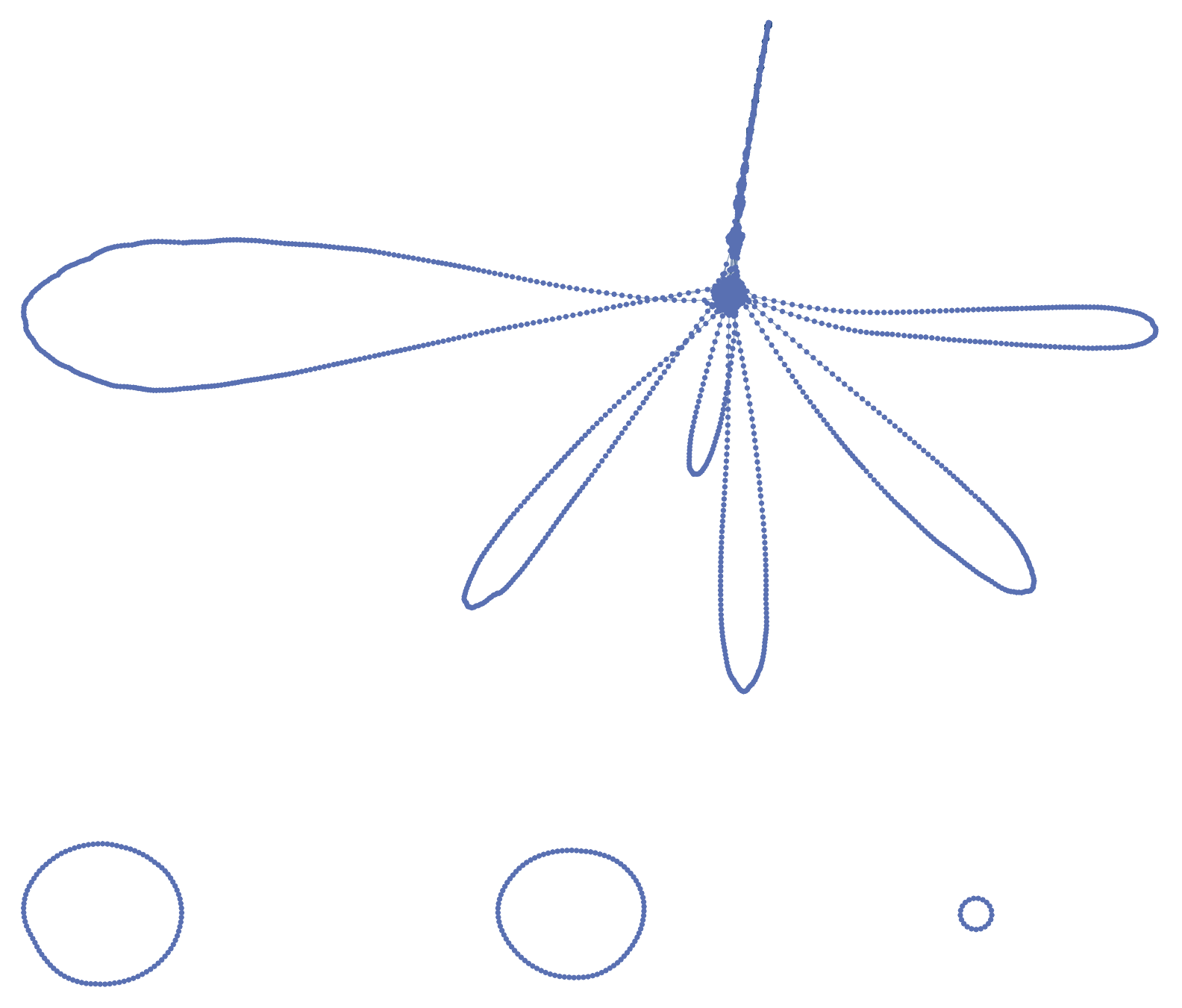}
\caption{An example of a maximally assortative network with $\gamma=3$, $k_{min}=2$. The minimum degree has been taken greater than 1 in this example in order to avoid the formation of isolated couples, which make up the large majority of maximally assortative networks if $k_{min}=1$. The length of the chains and their connection or disconnection to the the giant component have only a little effect on the total $r$ coefficient, and can therefore vary in each realization, even at very low temperature (here $T=10^{-6}$).
} 
\label{BA2}
\end{center}  
\end{figure}

When the rewiring temperature is very low one can actually observe that the value of $r$ changes by $10^{-5}$ or less, and the entropy becomes very small. This is because when the network is very close to its maximum possible assortativity or disassortativity, almost all rewiring steps are rejected, changes in $r$ are very small and the adjacency matrix remains practically constant (thus $\bar{A}_{ab}$ is either very close to 0 or 1, with small contribution to $S$).

\section{Results}

\subsection{Trajectories of assortative and disassortative rewiring in the $r$-$K$ plane at low $T$}
\label{results1}

\begin{figure}
\begin{center}
\includegraphics[width=14.4cm,height=8.4cm]{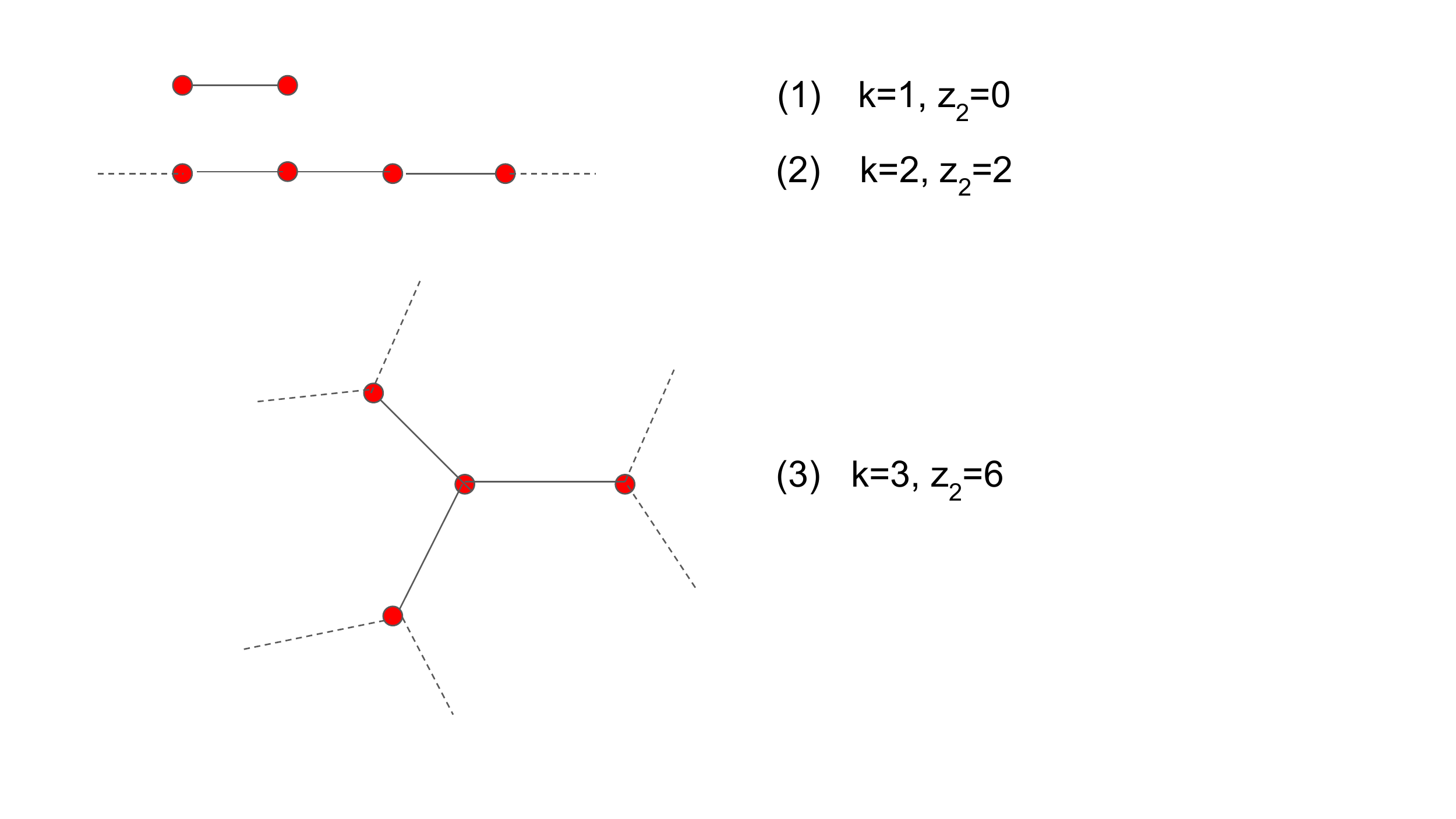}
\caption{Basic assortative building blocks classified through $k$ (node degree) and $z_2$ (number of second neighbors). Looking at the table which gives for each node of a strongly assortative network the number $z_2$ of its second neighbors, we find a large number of nodes with $k=1$, $z_2=0$ (corresponding to Graph (1), isolated couples), of nodes with $k=2$, $z_2=2$ (corresponding to Graph (2), chains) and of nodes with $k=3$, $z_2=6$ (corresponding to Graph (3), linked 3-stars).
} 
\label{blocks}
\end{center}  
\end{figure}

\begin{figure}
\begin{center}
\includegraphics[width=18cm,height=10.5cm]{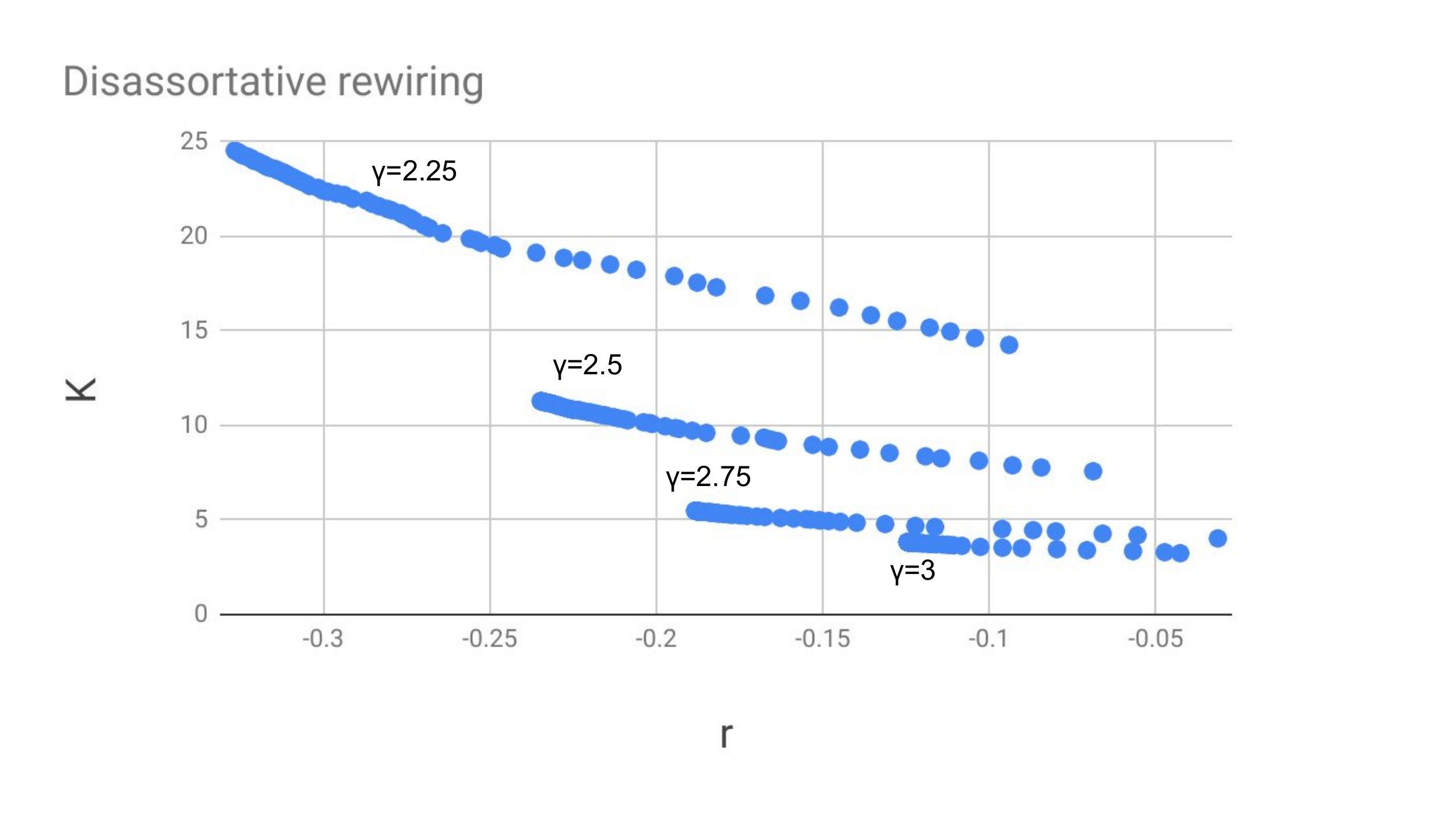}
\caption{Trajectories of disassortative rewiring in the $r$-$K$ plane. Each trajectory begins from the right (initial network obtained by uncorrelated rewiring) and converges to a maximally disassortative network on the left. Between two dots on the same trajectory there are 100 rewiring steps with return probability $\exp(\Delta r/T)$, $T=10^{-6}$. The number of nodes is $N=1500$.
} 
\label{traiettDIS}
\end{center}  
\end{figure}

Let us first describe the ``trajectories'' (see examples in Figs.\ \ref{traiettASS}, \ref{traiettDIS}) which are generated in the $r$-$K$ plane when we make a rewiring of the assortative kind, i.e., with rewiring step always accepted when $\Delta r>0$ and accepted with probability $\exp(-|\Delta r|/T)$ when $\Delta r<0$. The rewiring cycles described in this section are relatively short, in comparison to those of Sect.\ \ref{results2}, in which we need to reach equilibrium, and $T$ can be large.

The starting point of each trajectory is a network generated through a completely random rewiring. Note that especially for the lower values of $\gamma$ this ``configuration model'' network is not uncorrelated, but displays some structural disassortativity. 

We observe at the beginning a rapid increase of $r$ along the trajectory. For instance, for a network with 1500 nodes, in Fig.\ \ref{traiettASS} the data points represent the values of $r$ and $K$ during 80 sub-cycles of 100 rewirings each. The temperature $T$ is chosen to be quite low, compared to the magnitude order of the variation $\Delta r$. The latter is of the order of $10^{-3}$ as deduced from eq.\ (\ref{deltar}) and Tab.\ \ref{table2}. With these values and units, a temperature of $10^{-6}$ is sufficiently low to give a very small return probability in the Metropolis algorithm and to make the networks evolve quickly and  almost without fluctuations towards their maximum possible assortativity.

The points on the trajectory are seen to converge to a final spot, and typically if we perform 50 sub-cycles of 10000 steps (not shown in the figure), which take about one second to be completed, the values of $r$ and $K$ become constant up to the sixth decimal digit.
This value, however, depends on the degrees of the hubs effectively present in the network and is reproducible in different runs only if the hubs are generated with the method of cumulative probability (see Sect.\ \ref{alg}).

If at this stage we compute the Shannon entropy $S$ according to eq.\ (\ref{entropia}), we find $S=0$ exactly. This means that at this temperature the assortative rewiring causes the network in the long run to evolve towards a unique ``asymptotic'' state.

The maximum and minimum values (asymptotic values) that can be obtained for the $r$ coefficient of scale-free networks have been extensively studied in \cite{menche2010asymptotic}. Those results, however, are referred to networks whose minimum degree is relatively large, typically $k_{min}=6$; such networks are usually completely connected (giant component equal to 100\%). In this work we take instead $k_{min}=1$, because we are also interested into the network ``fragmentation'' caused by the strong assortativity and disassortativity. The giant component we find can be very small, down to 10\% or less. Most of the asymptotic networks consist, in the assortative case, of isolated couples made of nodes of degree $k=1$, and then of long chains with variable length, open or closed, made of nodes of degree $k=2$. Such structures can be easily identified in a complete graphical representation of the network or more ``economically'' in the table of the values of the number $z_2$ of second neighbors of each node. For instance, for each node of degree 1 belonging to a couple we obviously have $z_2=0$, and for each node of degree 2 belonging to a chain we have $z_2=2$ (except at the ends of the chain). Another less obvious occurrence observed in strongly disassortative networks is shown in Fig.\ \ref{blocks} and involves nodes of degree 3.

In the extreme assortative case the giant component contains all the major hubs, strongly connected among themselves.
From the strongly connected core depart long chains (see an example in Fig.\ \ref{BA2}) whose connection or disconnection to the core affects the size of giant component, but has very little influence on the value of $r$. In other words, a rewiring in which long chains are connected or disconnected at their ends can easily occur also at low temperature, with little effect on $r$ and on the entropy.

\begin{figure}
\begin{center}
\includegraphics[width=10cm,height=8.8cm]{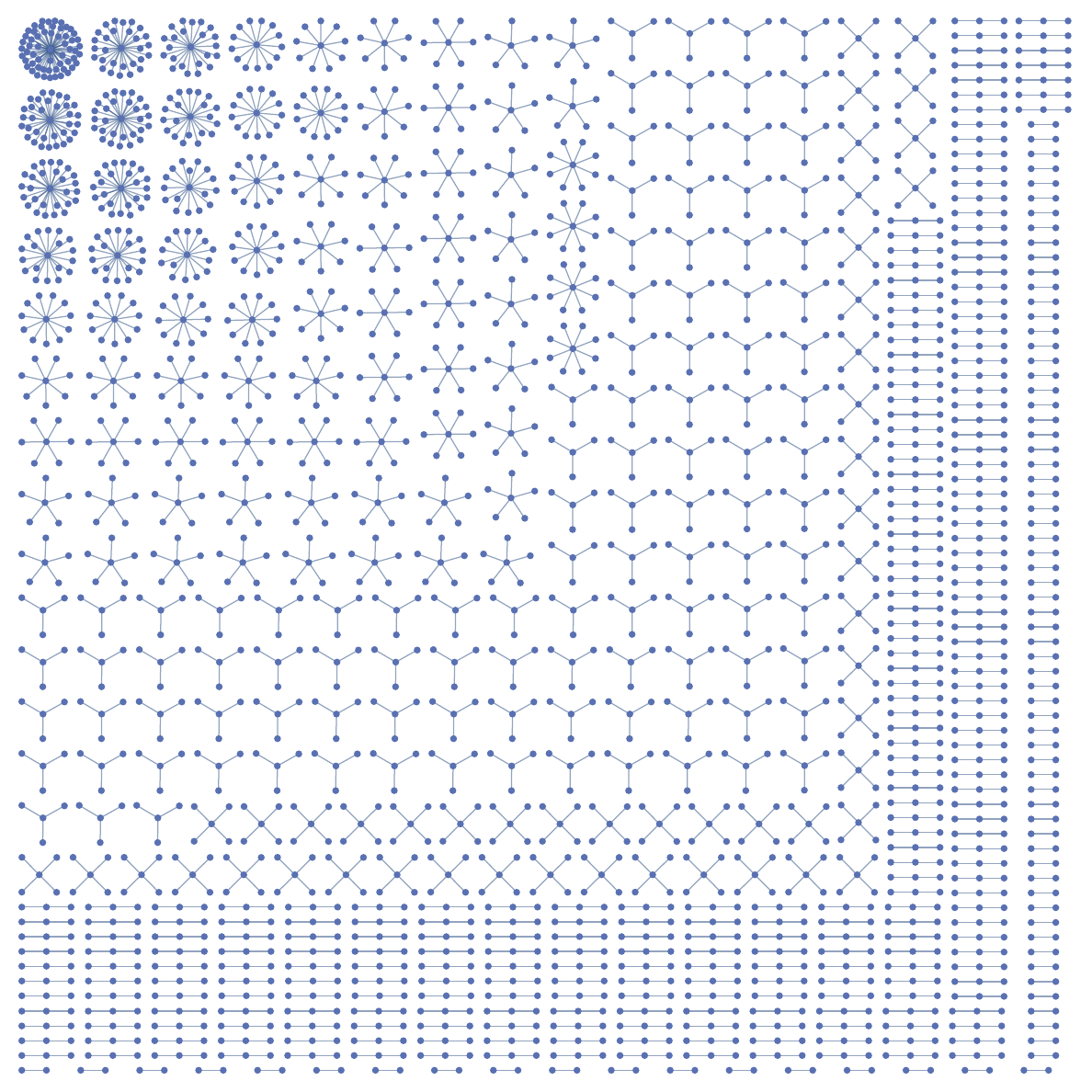}
\caption{Maximally disassortative network with $\gamma=2.75$, $k_{min}=1$. The network is completely fragmented into ``stars'' (compare \cite{moreno2003disease}). For all hubs the number of second neighbors is exactly zero.
} 
\label{super}
\end{center}  
\end{figure}

In the extreme disassortative case, especially for $\gamma$ close to 3, it may happen that the network is completely fragmented and there is no giant component. One example is shown in Fig.\ \ref{super}. This kind of networks has been studied theoretically in \cite{moreno2003disease}. Also in this case the table of values of the number $z_2$ of second neighbors is useful in order to identify some patterns without the need to visualize the entire network. For instance, all the hubs of the network in Fig.\ \ref{super} have $z_2=0$ exactly.

\subsection{Equilibrium rewiring at variable $T$}
\label{results2}

In this subsection we report some preliminary results from assortative rewiring cycles performed at variable temperature, in the range $10^{-5} \le T \le 10^{-2}$. These rewiring cycles consist of several long sub-cycles (typically 100 sub-cycles of $10^4$ rewiring steps), such that equilibrium is attained and entropy can be measured by averaging the adjacency matrix on each sub-cycle. We recall (see Sect.\ \ref{results1}) that the value of the temperature for the rewiring algorithm must be referred to the magnitude order of the variations of $r$, and this depends in turn on the number of nodes.

\begin{figure}
\begin{center}
\includegraphics[width=14.4cm,height=8.4cm]{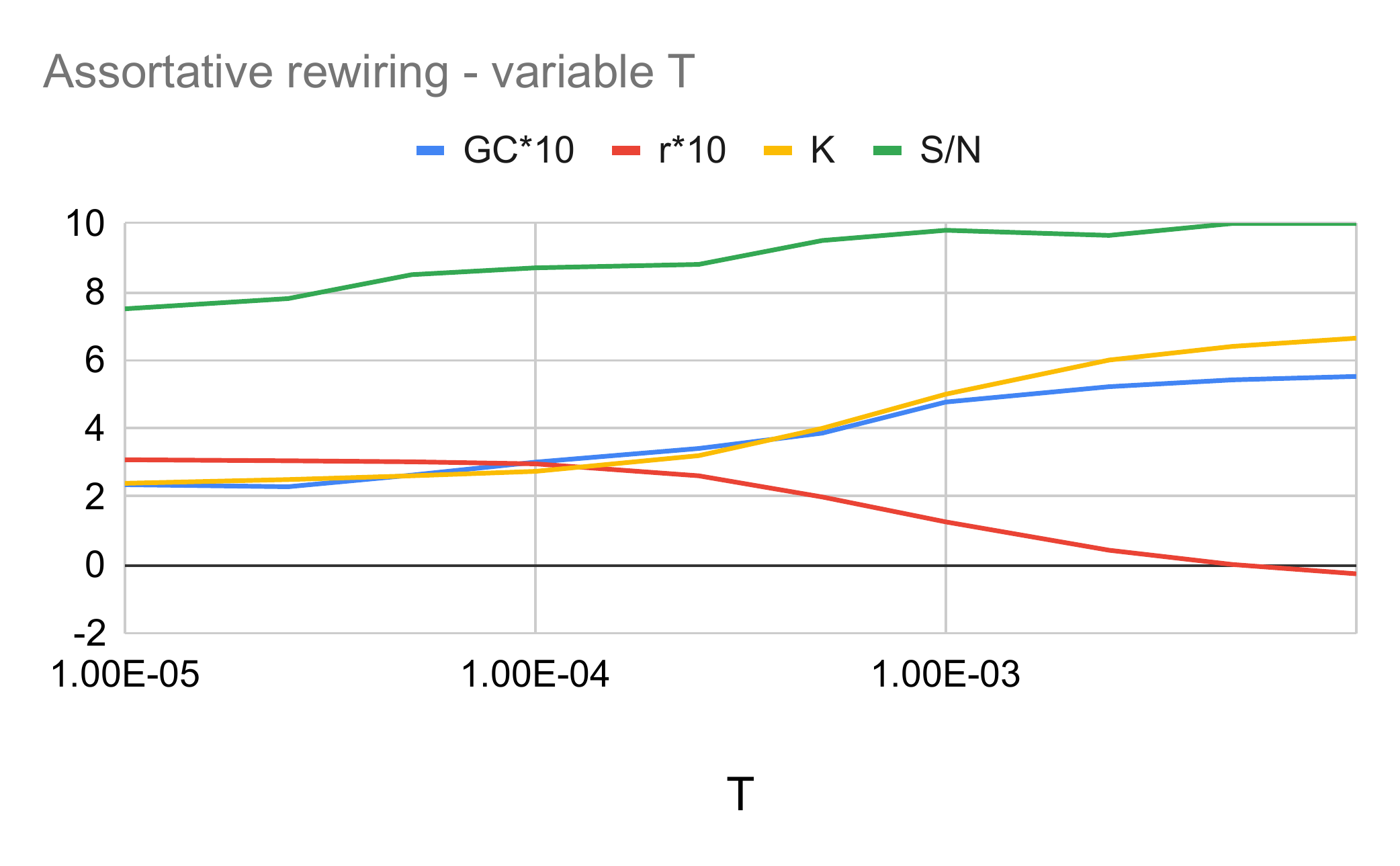}
\caption{Assortative rewiring in equilibrium at different temperatures $T$. (Network of 1000 nodes, with $\gamma=2.5$, $k_{min}=1$.) The entropy per node $S/N$ is shown, along with the average degree of the first neighbors $K$, the Newman assortativity coefficient $r$ multiplied by 10 (note the slight structural disassortativity at high temperature) and the fractional size of the giant component multiplied by 10.
} 
\label{var-T}
\end{center}  
\end{figure}

The data are still quite noisy and more statistics needs to be accumulated, for different values of $\gamma$, followed by extension to the case of disassortative rewiring. One of the objectives of these measurements is to verify the conjecture of \cite{johnson2010entropic} that disassortative networks are entropically favoured. Another interesting suggestion from the data in Fig.\ \ref{var-T} is that there is a correlation between $K$ and the size of the giant component, as $T$ varies. The exact anticorrelation between $r$ and $K$ according to eq.\ (\ref{deltaKbis}) is also evident.

\section{Conclusions}
\label{concls}

The method of assortative and disassortative rewiring at variable $T$ that we have presented in this work appears to be quite effective for the generation of correlated scale-free networks. At each step of the rewiring process, our algorithm permits a control of the variations of the assortativity coefficient $r$ and of the average degree of the nearest neighbors $K$. The two variations are actually connected through a general relation that we have proven in Sect.\ \ref{sezKvar2}.

We also have proven that the average number $\bar{z}_{2,B}$ of second neighbors in the branching approximation is constant in the rewiring. This property provides further evidence for the absence of epidemic threshold in scale-free networks with exponent $\gamma$ in the range $2<\gamma \le 3$.

If we represent an assortative or disassortative rewiring process at low temperature (i.e., with low return probability) in an $r$-$K$ plane, we obtain an almost linear trajectory converging towards a point which represents the maximally assortative or disassortative network having the given degree distribution. The position of the trajectory in the plane and its (negative) slope depend on the exponent $\gamma$. In general, the value of $K$ is smaller for assortative networks, compared to uncorrelated or disassortative networks having the same degree distribution. For a fixed value of $r$, $K$ is larger when $\gamma$ is smaller, therefore the trajectories with small $\gamma$ lie in the upper part of the plane.

The features of super-assortative and super-disassortative networks are found to depend quite strongly, for a given $\gamma$, on the minimum and maximum degree present in the network.

Preliminary evaluations of the network entropy, the size GC of the giant component, $K$ and $r$ as functions of the rewiring temperature confirm the exact anti-correlation between $r$ and $K$ and indicate a positive correlation between $K$ and GC.

\bibliographystyle{unsrt}
\bibliography{new3refsfornewman-2aprile}

\end{document}